\documentclass{PoS}

\usepackage{fleqn,epsf}

\def\mpi2{m_\pi^2}
\def\mK2{m_K^2}

\newcommand{\bea}{\begin{eqnarray}}
\newcommand{\eea}{\end{eqnarray}}
\newcommand{\be}{\begin{equation}}
\newcommand{\ee}{\end{equation}}

\def\lvec#1{\setbox0=\hbox{$#1$}
    \setbox1=\hbox{$\scriptstyle\leftarrow$}
    #1\kern-\wd0\smash{
    \raise\ht0\hbox{$\raise1pt\hbox{$\scriptstyle\leftarrow$}$}}
    \kern-\wd1\kern\wd0}
\def\rvec#1{\setbox0=\hbox{$#1$}
    \setbox1=\hbox{$\scriptstyle\rightarrow$}
    #1\kern-\wd0\smash{
    \raise\ht0\hbox{$\raise1pt\hbox{$\scriptstyle\rightarrow$}$}}
    \kern-\wd1\kern\wd0}

\newcommand{\VEV}[1]{\left\langle #1\right\rangle}
\newcommand{\Tr}{\mbox{Tr}}

\title{The QCD equation of state with asqtad staggered fermions }

\ShortTitle{The QCD equation of state with asqtad staggered fermions}

\author{C.~Bernard\\
        Physics Department, Washington University, St. Louis, MO 63130, USA}
\author{T.~Burch\\
        Institut f\"ur Theoretische Physik, Universit\"at Regensburg, D-93040 Regensburg, Germany}
\author{C.~DeTar\\
        Physics Department, University of Utah, Salt Lake City, UT 84112, USA}
\author{Steven~Gottlieb and \speaker{L.~Levkova}$^{\,}$\footnote{Current address: Physics
        Department, University of Utah, Salt Lake City, UT 84112, USA}\\
        Physics Department, Indiana University, Bloomington, IN 47405, USA}
\author{U.~M.~Heller\\
        American Physical Society, One Research Road, Box 9000, Ridge, NY 11961-9000, USA}
\author{J.~E.~Hetrick\\
        Physics Department, University of the Pacific, Stockton, CA 95211, USA}
\author{D.~B.~Renner and D.~Toussaint\\
        Physics Department, University of Arizona, Tucson, AZ 85721, USA} 
\author{R.~Sugar\\
        Physics Department, University of California, Santa Barbara, CA 93106, USA}

\abstract{
We report on our result for the equation of state (EOS) with a Symanzik improved
gauge action and the asqtad improved staggered fermion action at $N_t=4$ and
6. In our dynamical simulations with 2+1 flavors we use the inexact R algorithm
and here we estimate the finite step-size systematic error on the EOS.
Finally we discuss the non-zero chemical potential extension of the EOS and give
some preliminary results.}

\FullConference{XXIVth International Symposium on Lattice Field Theory\\
                 July 23-28, 2006\\
                 Tucson, Arizona, USA}

\begin{document}
\section{Introduction}
The determination of the equation of state (EOS) of strongly interacting matter
on the lattice requires actions with small discretization effects,
a realistic light quark spectrum and good control over the systematic errors in order  
for the result to be relevant in helping to understand experimental findings.
For our thermodynamics studies we use the asqtad quark action \cite{asq} for 2+1 flavors,
combined with a one-loop Symanzik improved gauge action \cite{sym}.
Both actions are highly improved and have discretization errors of $O(\alpha_sa^2,\, a^4)$ and $O(\alpha_s^2a^2,\, a^4)$, respectively.
We do our simulations along trajectories of constant physics using the dynamical R algorithm
\cite{Ralg} and thus our calculations are subject to finite step-size errors. Here we 
describe the method we use to estimate and correct for this systematic error.

To approximate the experimental conditions as closely as possible we
have started a project which extends our EOS calculation to a small 
non-zero chemical potential. For this purpose we use the 
Bielefeld--Swansea Taylor expansion method \cite{taylor} and preliminary results for the EOS at a set of
chemical potentials are presented.

\section{Simulation overview}
We study the quark-gluon system at a range of temperatures along trajectories of constant physics.
Such trajectories are defined by keeping the ratios $m_\pi/m_\rho$ 
and $m_{\eta_{ss}}/m_\phi$ fixed
while changing $a$ at a constant $N_t$. We work with two approximate 
trajectories; for both of them
the heavy (strange) quark mass $m_s$ is fixed to the physical 
value within about 20\%.
The light quark masses for the two trajectories 
are $m_{ud}\approx 0.2 m_s$ ($m_\pi/m_\rho\approx0.4$) and
$m_{ud}\approx 0.1 m_s$ ($m_\pi/m_\rho\approx0.3$), respectively. For both trajectories we have
performed calculations at $N_t =6$. In addition, a study at $N_t =4$ was done 
for the $m_{ud}\approx 0.1 m_s$ trajectory in order to compare 
the effect of the larger discretization errors on the EOS.
The constant physics trajectories are parameterized with  RG-inspired formulae 
using the hadron spectrum data at certain anchor points (see \cite{lat2005} 
for explicit formulae).

The lattice gauge configurations are generated with step sizes 
in the R algorithm chosen to be the smaller
of 0.02 and $2m_{ud}/3$, and occasionally smaller still.
Most of the gauge configurations used for the analysis were generated before
the RHMC algorithm~\cite{rhmc} became known.
Since Lattice 2005 \cite{lat2005}, we have doubled the statistics 
and added a new high-temperature run at $\beta=7.08$ ($a\approx0.086$ fm)
and a zero-temperature run at $\beta=6.275$ ($a\approx 0.232$ fm)
for the $N_t=4$ case.
For both trajectories new zero- and high-temperature runs at $\beta=6.85$ ($a\approx 0.110$ fm) 
were included as well. 
For the list of the rest of the run parameters see \cite{lat2005}. 
The lattice spacing $a$ has been determined at zero temperature
using the heavy quark potential ("$r_1$") with the overall scale set
by the $\Upsilon$ 1S-2S mass splitting~\cite{bot}.
We fit all the available data to an appropriate one-loop RG-inspired formula~\cite{lat2005} 
which gives the lattice spacing as a function of the quark masses and the gauge coupling
and allows us to determine the temperature along the constant physics trajectories.
\section{The EOS analytic form}
We determine the EOS using the integral method \cite{int}, where the pressure is calculated as an 
integral of the interaction measure and the energy density is a linear combination of both:
\bea
Ia^4&=&-6\frac{d \beta_{\rm pl}} {d \ln a}\Delta\VEV{P}
       -12 \frac{d \beta_{\rm rt}} {d \ln a}\Delta\VEV{R}
       -16 \frac{d \beta_{\rm ch}} {d \ln a}\Delta\VEV{C}\nonumber\label{eq:I}\\
    &&- \sum_f \frac{n_f}{4}\left[\frac{d (m_f a)}{d \ln a}
           \Delta\VEV{\bar \psi \psi}_f
       + \frac{d u_0}{d \ln a}
       \Delta\VEV{\bar\psi\, \frac{d M}{d u_0}\, \psi}_f\right],\\
pa^4 &=& -\int_{\ln a_0}^{\ln a} I(a^\prime)(a^{\prime})^4 d\ln a^\prime,\label{eq:p}\\
\varepsilon a^4 &= &Ia^4 +3pa^4.\label{eq:e}
\eea
For physics quantities and parameter definitions in the above see~\cite{lat2005}.
\section{Systematic errors and EOS results} 
Our calculation is affected by the following systematic errors: finite volume effects,
choice of the lower integration limit in Eq.~(\ref{eq:p}) and the finite step-size error
in the R algorithm. The first two errors are relatively straightforward to estimate.
By conducting an EOS calculation along the $m_{ud}\approx 0.1 m_s$, $N_t=4$ trajectory 
on a small ($8^3\times4$) volume and comparing it with the result of the calculation 
on a larger ($12^3\times4$)
volume, we have determined that the finite volume effects are negligible.
To estimate the error introduced by postulating the lower integration limit in Eq.~(\ref{eq:p})
to be at the coarsest available lattice scale 
in our simulations on a given trajectory, we have calculated the pressure of an ideal pion gas
at the corresponding lowest available temperature points. The value of the ideal pion gas pressure 
at these temperatures is about as large as the statistical errors we have. Thus, we have 
ignored this systematic error as well. 

\begin{figure}[h]
\begin{tabular}{ll}
 \epsfxsize=70mm
 \epsfbox{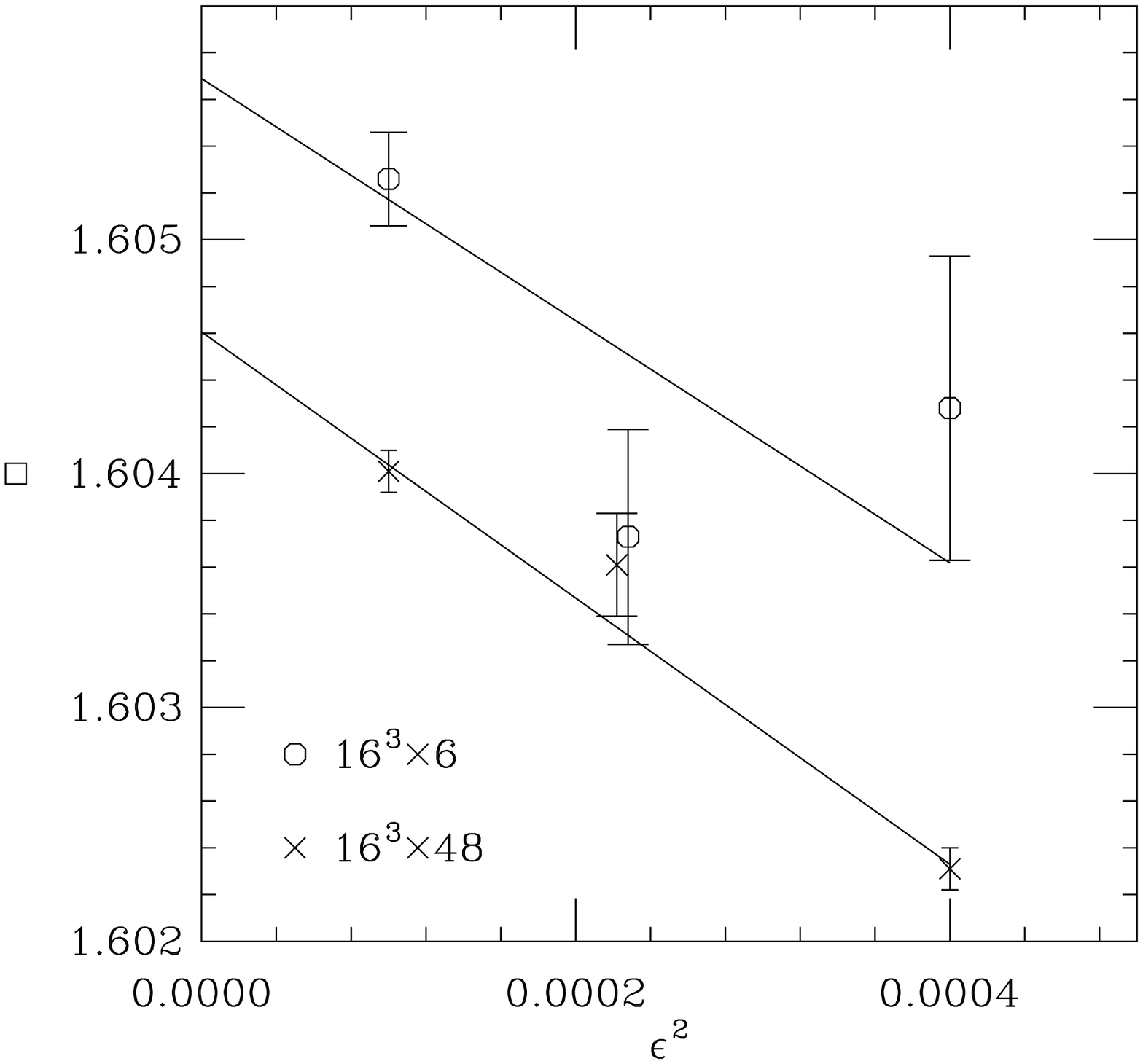}
&
 \epsfxsize=70mm
 \epsfbox{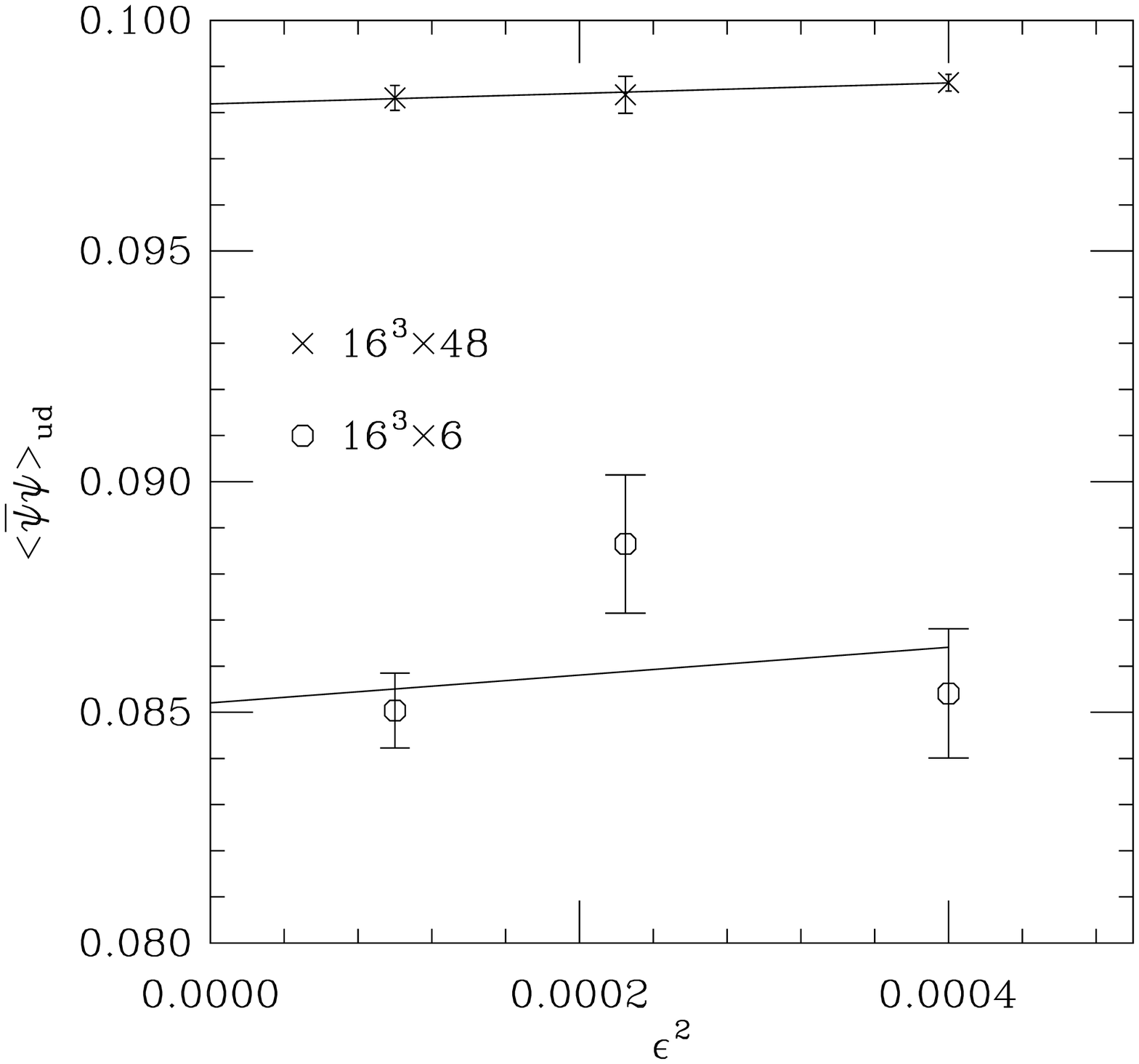}
\end{tabular}

\caption{Plaquette (left panel) and chiral condensate (right panel)
{\it vs} the squared step size $\epsilon^2$ for the improved action for
the ensemble at $\beta = 6.467$, $am_{ud} = 0.01676$ and $am_s =
0.0821$.  The squared step size used for production of this ensemble is 0.0001.}
\label{fig:obsvseps}
\end{figure}

\begin{figure}[h]
\begin{tabular}{ll}
 \epsfxsize=70mm
 \epsfbox{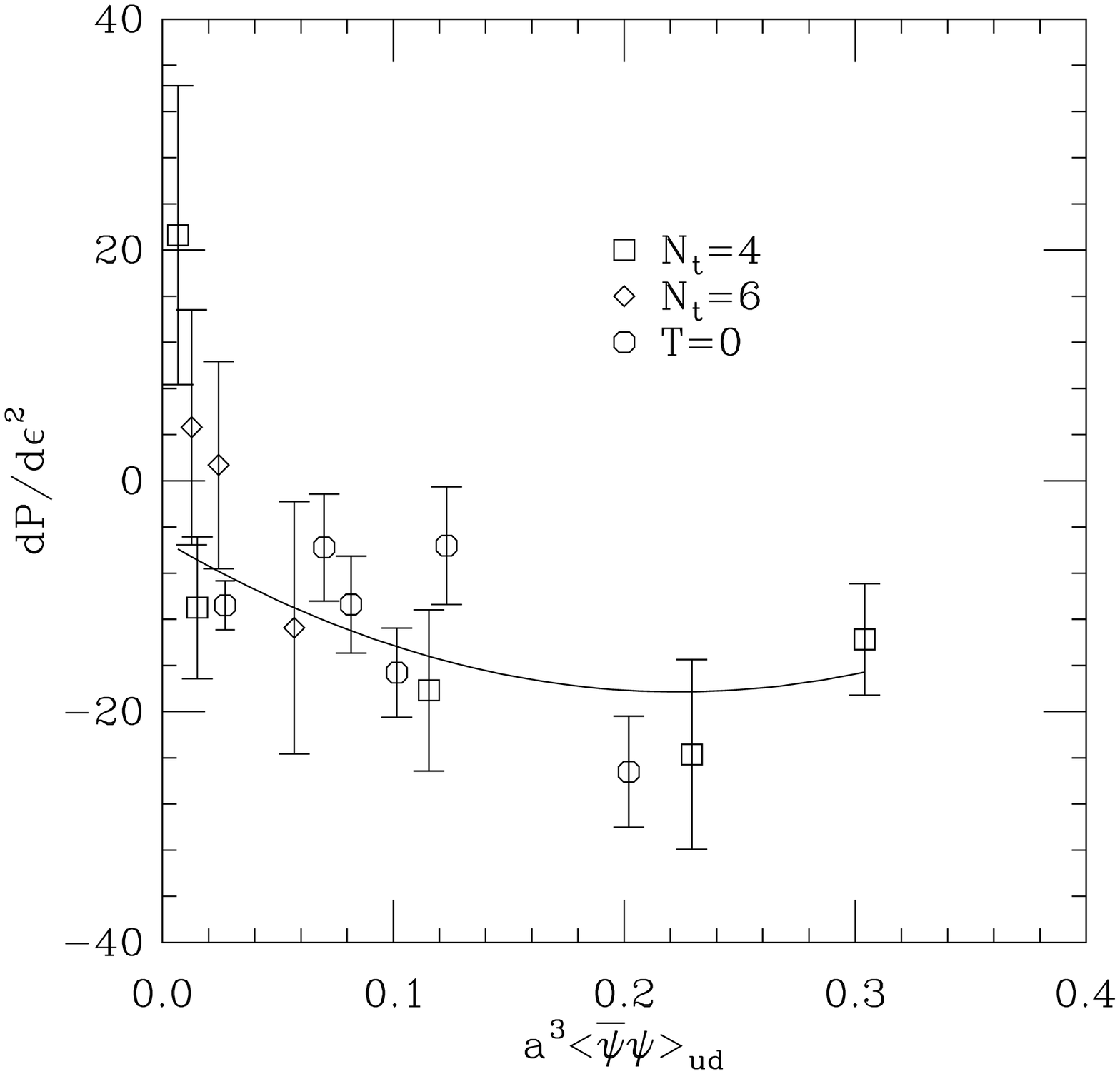}
&
 \epsfxsize=70mm
 \epsfbox{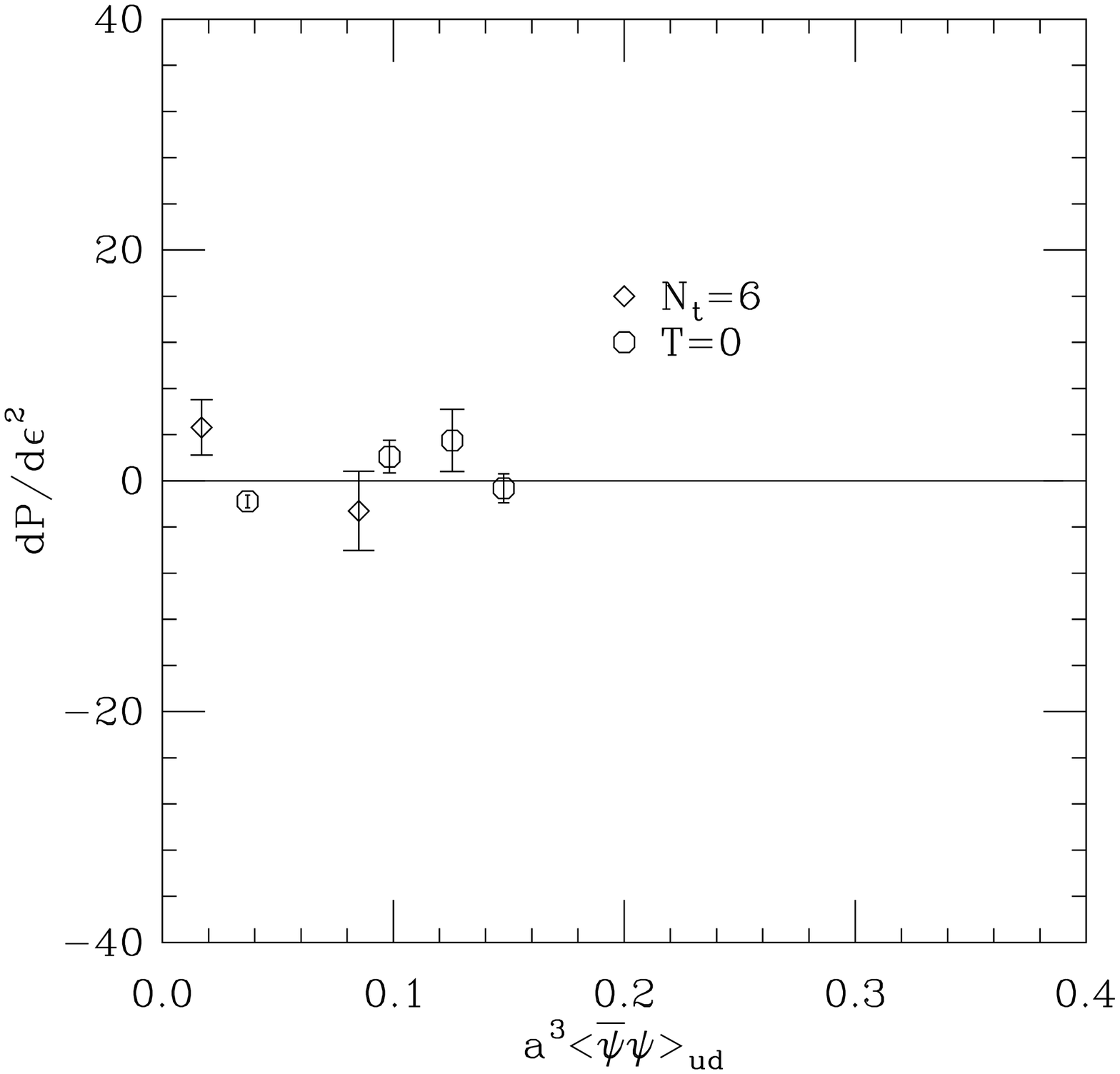}
\end{tabular}
\caption{Plaquette slopes for the $m_{ud}\approx 0.1 m_s$ trajectory (left) and the
$m_{ud}\approx 0.2 m_s$ trajectory (right) {\it vs.} the chiral condensate.}
\label{fig:slope_pbp}
\end{figure}

\begin{figure}[h]
\epsfxsize=80mm
\begin{center} 
\epsfbox{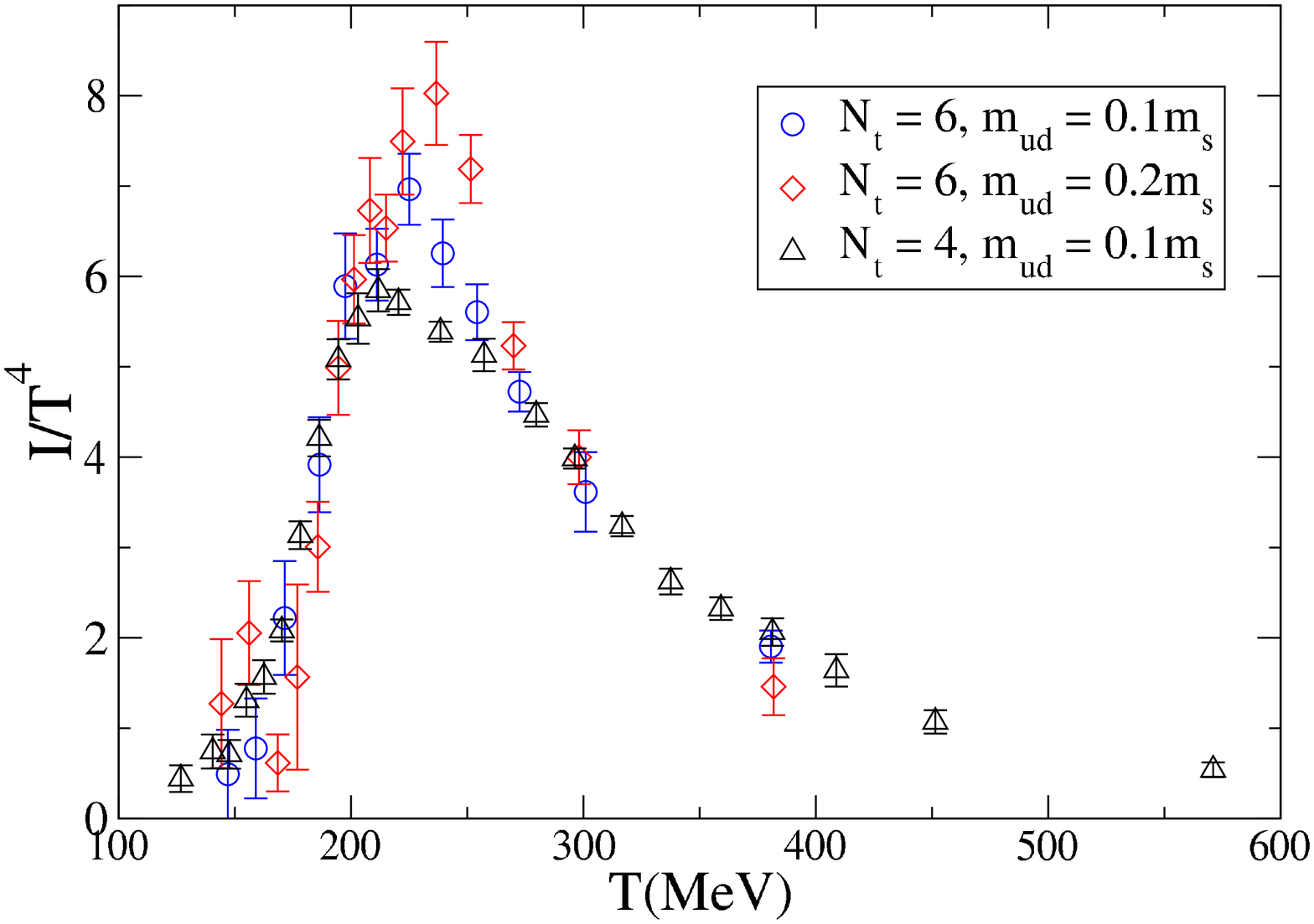}
\end{center}\vspace{-0.3cm} 
\caption{Interaction measure {\it vs.} temperature for both constant physics 
trajectories and $N_t$'s.}
\label{fig:I}
\end{figure}

\begin{figure}[h]
\begin{tabular}{ll}
 \epsfxsize=72mm
 \epsfbox{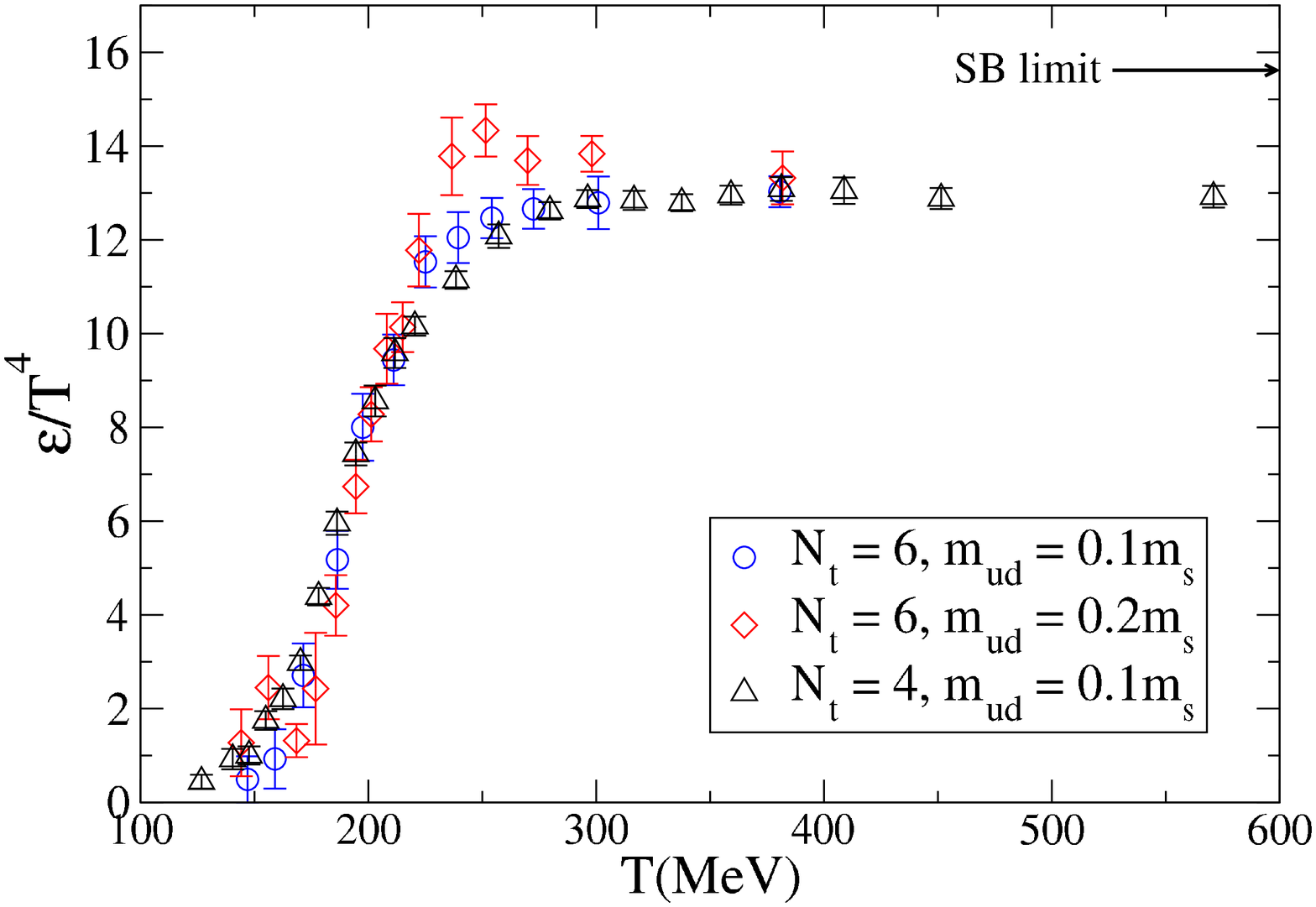}
&
 \epsfxsize=72mm
 \epsfbox{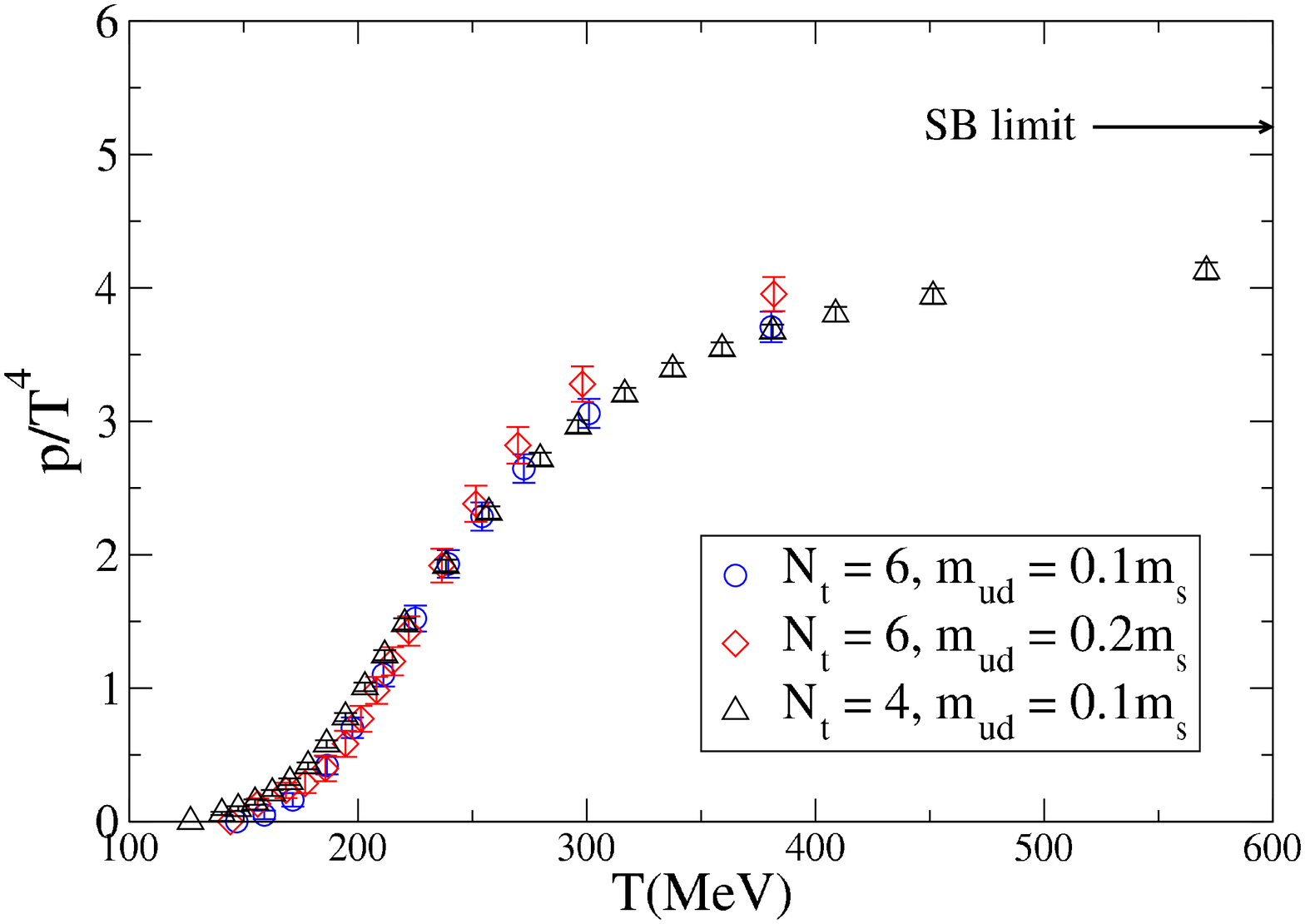}
\end{tabular}
\caption{Energy density (left) and pressure (right) {\it vs.} temperature 
for both constant physics trajectories and $N_t$'s.}
\label{fig:EP}
\end{figure}

The finite step-size error determination is a much more involved procedure. We have performed
some additional simulations at different step sizes using the R algorithm and, recently, some using
the exact RHMC algorithm. We measured most of the gluonic and the fermionic observables
in Eq.~(\ref{eq:I}) at different step sizes.
Figure~\ref{fig:obsvseps} shows the step-size dependence of the plaquette 
and the light chiral condensate for one case. 
The step-size error in the plaquette variable has a potentially significant effect on the EOS,
whereas the step-size error in the chiral condensate (and the rest of the fermionic observables)
is negligible. 
We have found that the gluonic observables (plaquette, rectangle and parallelogram) 
have similar step-size dependence slopes. For this reason we have concentrated on studying 
the finite step-size corrections mainly for the plaquettes and applied the same 
corrections to the rest of the gluonic observables.
Figure~\ref{fig:slope_pbp} shows the plaquette
slopes determined from a set of runs with different step sizes for both trajectories.
We see that for the  $m_{ud}\approx 0.2 m_s$ trajectory the slopes are small which means 
that the finite step-size errors can be ignored in this case. For the $m_{ud}\approx 0.1 m_s$
trajectory we have more data for the plaquette slopes from both $N_t=6$ and 4 simulations. We fit our data
to a quadratic form and use it to correct the EOS. Still, even in this case the finite step-size 
errors are quite small. The finite step-size corrections which we showed at the Lattice 
2006 conference were overestimated due to the limited amount of data on step-size dependence
we had at that time. 
Figure~\ref{fig:I} and \ref{fig:EP} show our results for the interaction measure, 
pressure and energy density after the small finite step-size corrections in the
$m_{ud}\approx 0.1 m_s$ trajectory have been included. 
\section{The EOS with non-zero chemical potential}
\begin{figure}[h]
\epsfxsize=\hsize
\begin{center} 
\epsfbox{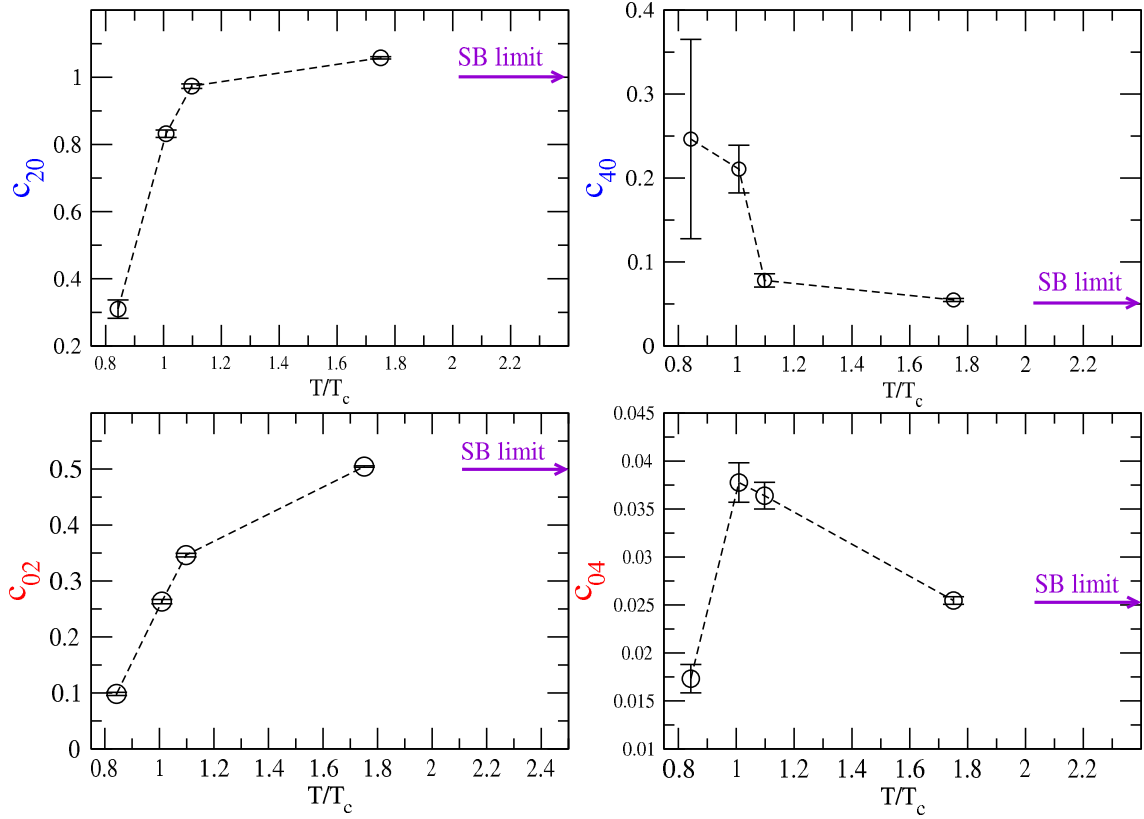}
\end{center}\vspace{-0.3cm} 
\caption{Some Taylor coefficients in the pressure expansion for the 
$m_{ud}\approx 0.1 m_s$, $N_t=4$ trajectory.}
\label{fig:Cnm}
\end{figure}

\begin{figure}[h]
\begin{tabular}{ll}
 \epsfxsize=72mm
 \epsfbox{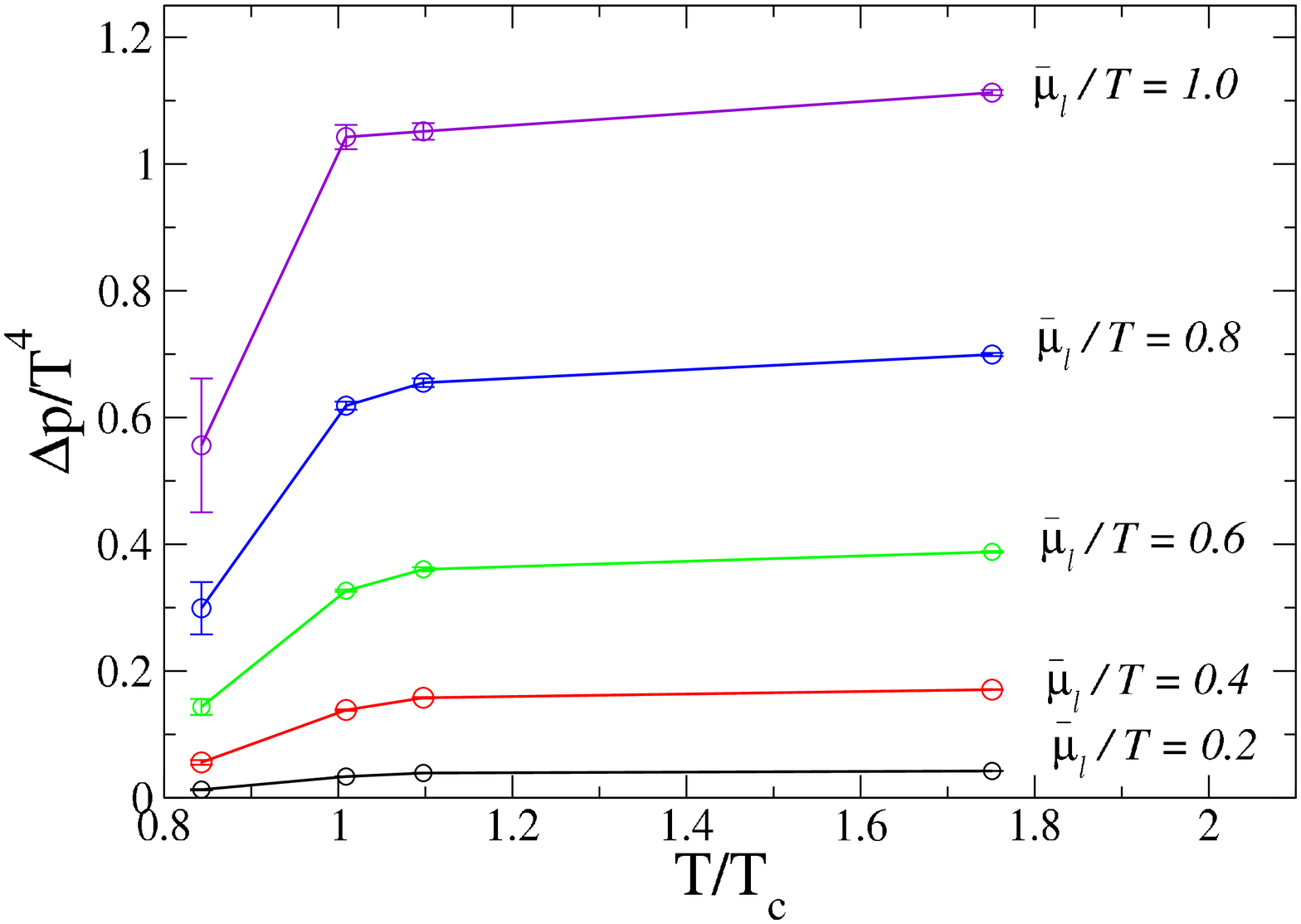}
&
 \epsfxsize=72mm
 \epsfbox{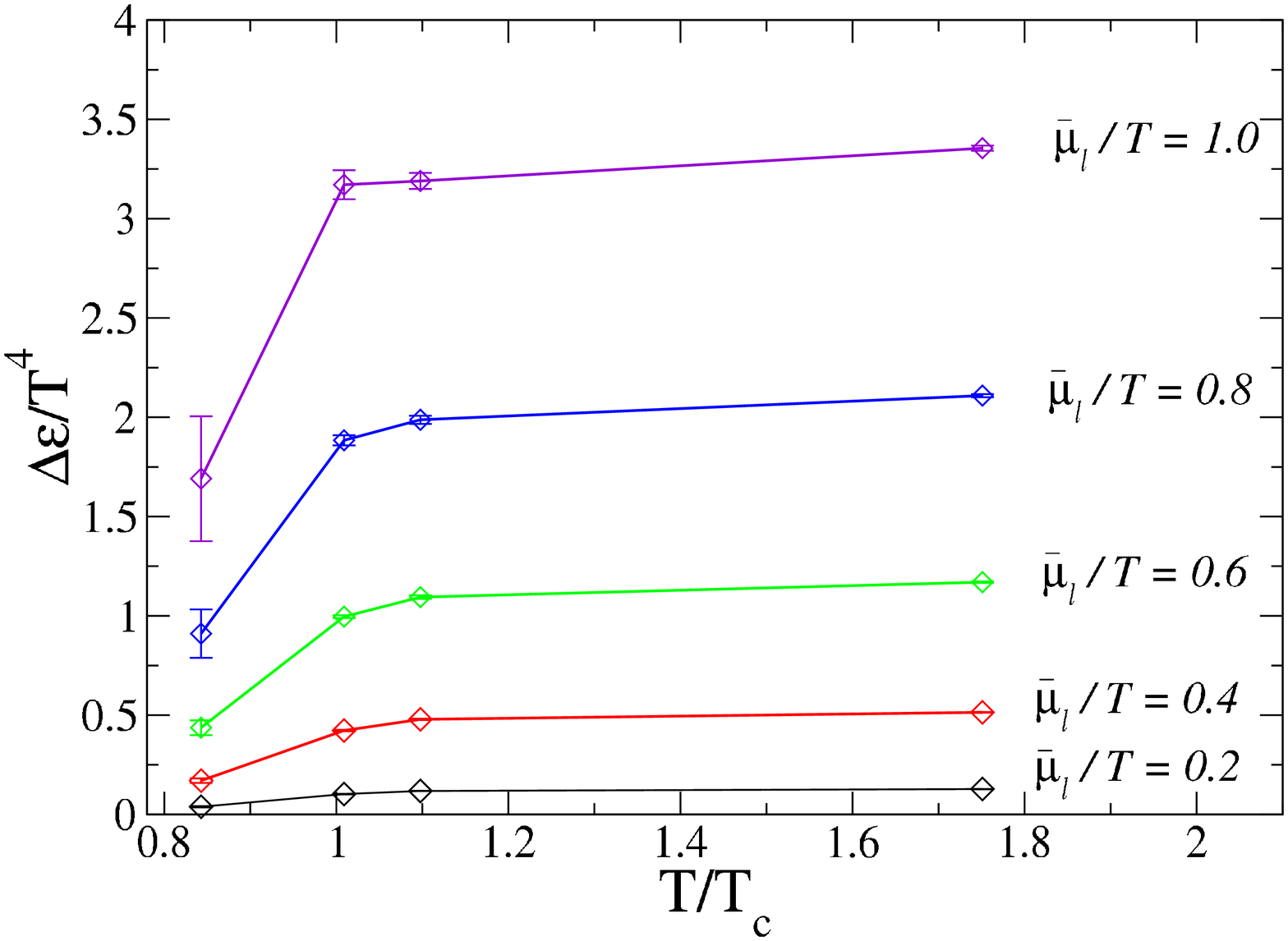}
\end{tabular}
\caption{Pressure correction $\Delta p = p(\mu_l) - p(\mu_l=0)$ 
and energy density correction $\Delta \varepsilon = \varepsilon(\mu_l) - \varepsilon(\mu_l=0)$ for the
$m_{ud}\approx 0.1 m_s$, $N_t=4$ trajectory to $O(\mu^4)$ and at $\mu_h=0$.}
\label{fig:EOSmu}
\end{figure}

To include a non-zero chemical potential in the EOS calculation we use the 
Bielefeld--Swansea Taylor expansion method \cite{taylor}.
According to this method the pressure can be expanded as:
\be
\frac{p}{T^4}=\frac{\ln Z} {VT^3}=
\sum_{n,m=0}^\infty c_{nm}(T) \left(\frac{\bar{\mu}_l}{T}\right)^n
\left(\frac{\bar{\mu}_h}{T}\right)^m,\label{eq:pmu}
\ee
where $Z$ is the partition function, and $\bar{\mu}_{l,h}$ are the
chemical potentials for the light and heavy quarks, respectively.
We note that the problems of rooted staggered fermions at non-zero
chemical potential \cite{Golterman:2006rw} are not relevant here since all
the expansion coefficients are evaluated in the $\mu_{l,h}=0$ theory.
Due to CP symmetry the non-zero terms in the series have even $n+m$. The non-zero
coefficients in the above are
\be
c_{nm} (T)=
\frac{1}{n!}\frac{1}{m!}\frac{N_t^{3}}{ N_s^3}\frac{\partial^{n+m}\ln Z}
{\partial(\mu_l N_t)^n\partial(\mu_h N_t)^m}\biggr\vert_{\mu_{l,h}=0} \quad,
\label{eq:cn}
\ee
where now the $\mu_{l,h}$ are the chemical potentials in lattice units. 
Similarly, for the interaction measure we have:
\be
{I\over T^4}=-{N_t^3\over N_s^3}{d\ln Z \over d\ln a}=\sum_{n,m}^\infty b_{nm}(T)
\left({\bar{\mu}_l\over T}\right)^n
\left({\bar{\mu}_h\over T}\right)^m,
\ee
where again only terms with even $n+m$ are non-zero and
\be
b_{nm}(T) = \left.-{1\over n!m!}{N_t^3\over N_s^3}{\partial^{n+m} \over 
\partial(\mu_l N_t)^n\partial(\mu_h N_t)^m}
\right|_{\mu_{l,h}=0}\left({d\ln Z\over d\ln a}
\right).
\ee
To determine the EOS, we need to calculate derivatives of the asqtad fermion matrix such as
\be
\frac{\partial^{n} \ln \det M_{l,h}}{\partial \mu_{l,h}^n}, \hspace{1cm}
\frac{\partial^{n} \,\Tr M_{l,h}^{-1}}{\partial \mu_{l,h}^n}, \hspace{1cm}
\frac{\partial^{n} \,\Tr (M_{l,h}^{-1}\frac{dM_{l,h}}{du_0})}{\partial \mu_{l,h}^n}.
\ee
These derivatives are estimated on the ensembles of lattices along a
constant physics trajectory using 200 random sources in the region 
of the phase transition/crossover and 100 sources outside that region. We intend to study the EOS
up to $O(\mu^6)$ on both constant physics trajectories. Currently we have data for only one of them
and  not enough statistics to resolve the sixth-order terms. 
Thus the preliminary results for the EOS we present here are calculated to
$O(\mu^4)$ on the $m_{ud}\approx 0.1 m_s$, $N_t=4$ trajectory only. We also set $\mu_h=0$ here. 
Figure~\ref{fig:Cnm} shows our result for some of the coefficients
involved in Eq.~(\ref{eq:pmu}). An interesting feature is that the coefficients
quickly reach the continuum Stefan-Boltzmann limit above $T_c$.
The pressure and energy density corrections due to the non-zero chemical potential
are shown in Figure~\ref{fig:EOSmu}.

\section{Conclusions}
We have calculated the EOS along two constant physics trajectories with $m_{ud}\approx 0.1m_s$
and $0.2m_s$ using the R algorithm. To estimate the finite step-size errors we have 
performed a number of additional R algorithm simulations at different step-sizes 
and a few RHMC simulations along the constant physics trajectories. The analysis of this additional 
data shows that the finite step-size errors are negligible on the $m_{ud}\approx 0.2m_s$
trajectory and small (in most cases less a than few percent) on the  $m_{ud}\approx 0.1m_s$ trajectory.

We have started a non-zero chemical potential study of the EOS with 2+1 flavors using the 
Bielefeld--Swansea Taylor expansion method and presented preliminary EOS results for
a number of different $\bar{\mu}_l/T$ values.

\begin{center} 
{\bf Acknowledgments}\\
\end{center}
We thank Maarten Golterman for discussions.
This work was supported by the US DOE and NSF. 
Computations were performed at CHPC (Utah), FNAL, FSU, IU, NCSA and UCSB.

\end{document}